\def\degree{\hbox{$^\circ$}}

\documentclass[review]{elsarticle}

\usepackage{amssymb}

\usepackage{graphicx}
\usepackage{hyperref}
\usepackage{amssymb,amsmath}
\usepackage{makecell}

\hypersetup{pdfauthor={Name}}

\journal{Chaos, Solitons \& Fractals}

\begin{document}

\begin{frontmatter}



\title{Increased earthquake rate prior to mainshocks}


\author[BIU]{Eitan E. Asher}
\author[BIU]{Shlomo Havlin}
\author[BIU,NRCN]{Shay Moshel}
\author[BIDR]{Yosef Ashkenazy\corref{cor1}}
\ead{ashkena@bgu.ac.il}
\cortext[cor1]{corresponding author}
\address[BIU]{Department of Physics, Bar-Ilan University, Ramat Gan, Israel}
\address[NRCN]{Nuclear Research Center Negev, 84190, Beer-Sheba, Israel}
\address[BIDR]{Department of Environmental Physics, BIDR, Ben-Gurion University, Midreshet Ben-Gurion, Israel}

\begin{abstract}
  According to the Omori-Utsu law, the rate of aftershocks after a mainshock decays as a power law with an exponent close to 1. This well-established law was intensively used in the past to study and model the statistical properties of earthquakes. Moreover, according to the so-called inverse Omori law, the rate of earthquakes should also increase prior to a mainshock---this law has received much less attention due to its large uncertainty. Here, we mainly study the inverse Omori law based on a highly detailed Southern California earthquake catalog, which is complete for magnitudes $m_c\geq1$ or even lower. First, we develop a technique to identify mainshocks, foreshocks, and aftershocks. We then find, based on a statistical procedure we developed, that the rate of earthquakes is higher a few days prior to a mainshock. We find that this increase is much smaller for (a) a catalog with a magnitude threshold of $m_c = 2.5$ and (b) for the Epidemic-Type Aftershocks Sequence (ETAS) model catalogs, even when used with a small magnitude threshold (i.e., $m_c = 1$). We also analyze the rate of aftershocks after mainshocks and find that the Omori-Utsu law does not hold for many individual mainshocks and that it may be valid only statistically when considering many mainshocks together. Yet, the analysis of the ETAS model based on the Omori-Utsu law exhibits similar behavior as that of the real catalogs, indicating the validity of this law.
\end{abstract}


\begin{highlights}
\item We developed and implemented an algorithm to identify mainshocks, aftershocks, and foreshocks.
\item We found that the earthquake rate is significantly higher than the average rate prior to mainshocks.
\item The increased earthquake rate prior to mainshocks is evident in detailed earthquake catalogs.
\item The ETAS model does not reproduce the increased earthquake rate before mainshocks.
\end{highlights}


\end{frontmatter}


\section{Introduction}
Earthquakes pose a serious threat to human life and property. There are countless examples of the devastating power of major earthquakes, and thus, earthquake mitigation efforts have been intensive. The use of modern building standards and social strategies has been adopted to minimize earthquake damage, especially in tectonically active areas. A reliable method of predicting earthquakes, using an observable signal, has not yet been developed~\cite[][]{jordan2006earthquake, jordan2011operational, ogata2017statistics}. The current earthquake predictability, which is of low effectiveness, is based on known empirical seismic laws.

An active direction for earthquake forecasting is using statistical models, such as the Epidemic-Type Aftershocks Sequence (ETAS) model~\cite[][]{ogata1988statistical,zhang2021improved}, which aims to determine the aftershock rate after a mainshock. Such models are based on empirical earthquake laws, including the Gutenberg-Richter law \cite[][]{gutenberg1944frequency,gutenberg1956earthquake} and the Omori-Utsu law \cite[][]{omori1894after,utsu1961statistical}. These models, however, are also not capable of forecasting mainshocks \cite[][]{ogata1988statistical,ogata1998space,gerstenberger2005real}, and their aftershock prediction success is limited \cite[][]{woessner2011retrospective, taroni2018prospective}. It is clear that even a small improvement in the forecasting skill of major earthquakes and aftershocks is urgently needed.

The Omori-Utsu law \cite[][]{omori1894after} is an empirical law that determines the rate of events (aftershocks) after a mainshock. According to this law, the temporal decay of the aftershock rate is expressed as 
\begin{equation}
    r(t) = \frac{k}{(c+t)^p}
\end{equation}
where $t$ is the time after a mainshock, $c$ is a case-dependent time scale, $k$ is a productivity parameter that depends on the magnitude of the mainshock, and the power $p$ is a parameter that quantifies the decay rate where typically it is assumed that $p\approx1$. According to the Gutenberg-Richter law, the distribution of earthquake magnitudes decays exponentially. A generalized version of the Omori-Utsu law that combines the known statistical laws for the earthquakes of Båth, Gutenberg-Richter, and the Omori-Utsu has been proposed \cite[][]{shcherbakov2004generalized} and tested by \cite{davidsen2015generalized} on four strong mainshocks in Southern California. Here, we will focus on the standard version of this law.

A less familiar version of the Omori law is the inverse Omori law
\cite[][]{ogata2017statistics, papazachos1973time, kagan1978statistical, jones1979some, console1993foreshock} which states that the rate of earthquakes increases prior to mainshocks, typically, within a fault size distance from the epicenter ($\sim$100 km).  Those events are determined a posteriori (as foreshocks) and are also referred to as an anomalous foreshock sequence. In principle, such behavior could be used for forecasting strong earthquakes.

There is still a controversy regarding the reason for higher event rates before mainshocks. Two debated conceptual hypotheses about earthquake nucleation, the ``pre-slip model'' and the ``cascade model,'' offer insight into the role played by foreshocks in earthquake predictability. In the former model, foreshocks are triggered by an aseismic slip anticipating the mainshock, while in the latter model, foreshocks are like any other earthquake that triggers one another to ultimately become larger (the mainshock). However, despite remarkable achievements with regard to foreshock rate \cite[][]{chen2016analysis,moutote2021rare}, an increased earthquake rate prior to large earthquakes has not always been found, limiting its use for forecasting.

Recently, \cite{trugman2019pervasive} utilized the comprehensive catalog of \cite{ross2019searching} to examine the earthquake rate preceding mainshocks. They posited a null hypothesis suggesting that the earthquake rate before mainshocks should correspond to the background rate, characterized by gamma-distributed interevent intervals. Their findings revealed a notable increase in earthquake rate 10-20 days prior to the mainshock within a 10 km radius of the epicenter; we further elaborate on this study in the conclusions section.
In addition to the above, a recent study by \cite{manganiello2021anomalous} showed that high heat flow regions in Southern California are associated with elevated foreshock rates. However, since it is typically correlated with small mainshock magnitudes, it is less useful for forecasting major events. Here, we aim to revisit this observation based on a high-resolution earthquake catalog. Such a detailed catalog has recently been constructed based on machine learning algorithms \cite[][]{ross2019searching, xie2020promise,li2018machine,beroza2021machine, tan2021machine}. Such earthquake catalogs may improve earthquake forecasting; we aim to test the validity of the inverse Omori law and the conventional Omori-Utsu law based on a detailed earthquake catalog and to determine whether such catalogs can improve the forecasting of impending earthquakes.

There are two main goals in our study: a) to study the rate of earthquakes (foreshocks) before mainshocks, and b) to study the rate of earthquakes (aftershocks) after a mainshock to identify the range of validity of the Omori-Utsu law. For this purpose, we use and compare both the detailed and less detailed catalogs of the Southern California region. We first develop a new technique to identify mainshocks and foreshocks. Then, we determine whether the number of events in a fixed time window, $D_{\mathrm{fs}}$, is significantly higher than the corresponding number in surrogate earthquake catalogs. We also test whether the higher rate found in the data is similar to or higher than that of the ETAS model. In addition, we test the validity of the Omori-Utsu law for aftershocks, both in real catalogs and ETAS model catalogs. Our results suggest that the Omori-Utsu law may be valid statistically, but does not hold for most individual mainshocks.

\begin{figure}[th!]
\centering
	\includegraphics[width=0.9\linewidth]{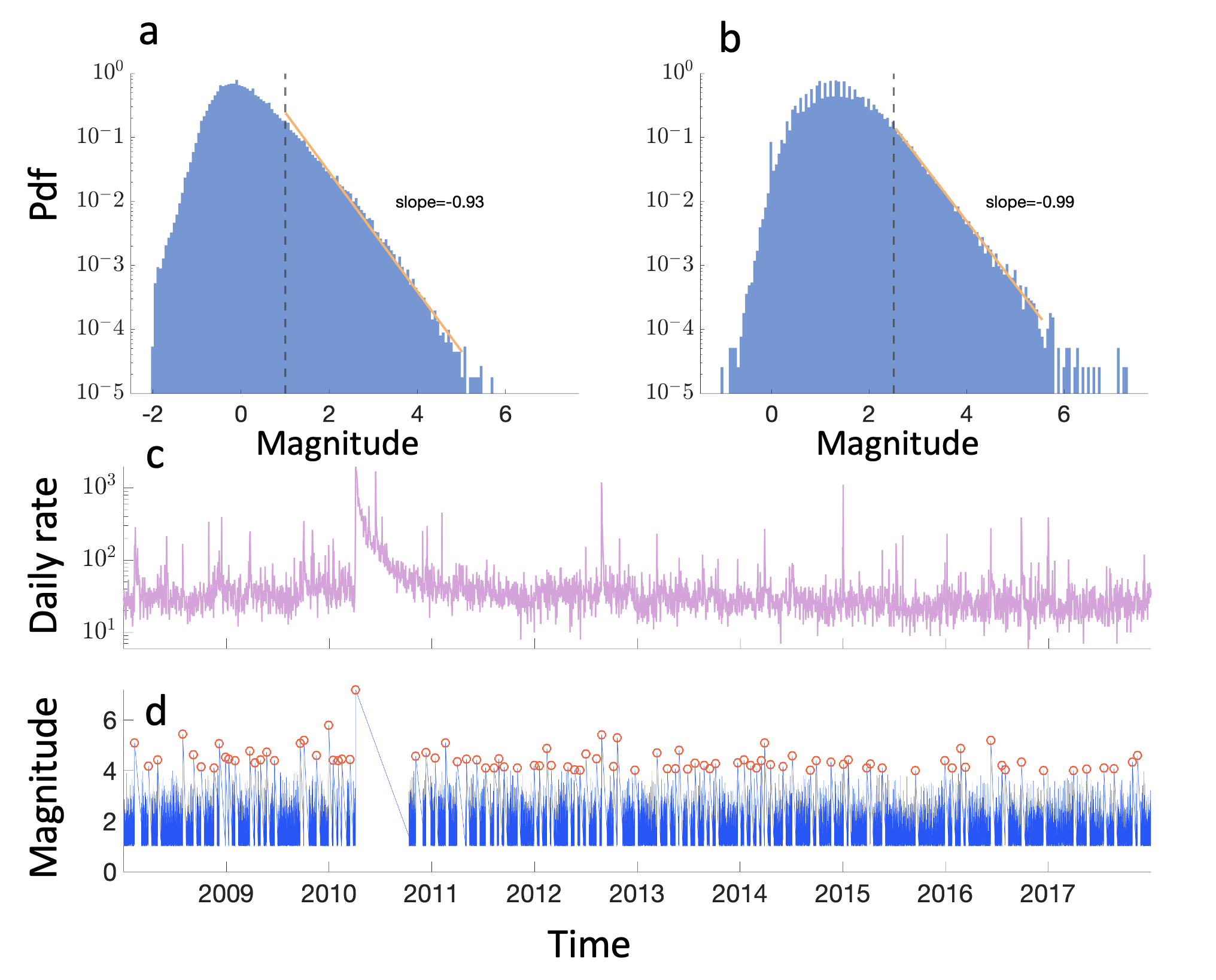}
	\caption{ {\bf Magnitude distribution, daily earthquake rate, and mainshock identification.} Magnitude distributions of \textbf{a}.  the Ross { et al.} \cite{ross2019searching} and \textbf{b}.  the Hauksson { et al.} \cite{hauksson2012waveform} catalogs. The Ross { et al.} and Hauksson { et al.} catalogs are complete for earthquakes of magnitudes larger than or equal to 0.3 and 2.5, respectively. The slope of the tail of the distribution is shown, and as expected, it is around -1; the slope was calculated over the range of magnitudes between 1 and 5 for the Ross { et al.} catalog and between 2.5 and 5.5 for the Hauksson { et al.} catalog. \textbf{c}. The daily earthquake rate of the Ross { et al.} catalog used in this study where the magnitude threshold is 1. \textbf{d}. Earthquake magnitudes (above the magnitude threshold 1) of the Ross { et al.} catalog versus time, after excluding the aftershocks using our proposed identification algorithm. Mainshocks are marked with red circles. As an example, consider the El Mayor-Cucapah earthquake, a 7.2 magnitude event that occurred on April 4, 2010. This event was followed by seven months of daily rate decline until the earthquake rate returned to the mean earthquake rate. This period is the relaxation time period, and the events in this period have been removed since they were considered aftershocks. See the Methodology and Data section for more details. }
	\label{fig:Fig1}
\end{figure}

\section{Methodology and Data}
\subsection{Determining a mainshock and defining the aftershock period}

We first develop an iterative method to identify mainshocks based on their magnitude and the earthquake rate following large events. The specific steps are described below. We consider earthquake events with magnitudes larger than or equal to a minimum magnitude threshold, $m_c$; the event dates are marked with $(d_1,d_2, ..., d_k)$, with their corresponding event magnitudes $(m_1,m_2, ..., m_k)$, where $k$ is the number of events above the magnitude threshold, based on the entire catalog. Next, we calculate the average daily event rate, $\langle r_d \rangle = \frac{k}{T}$ where $T = d_k-d_1$ is the time span (in days) of the catalog.

\begin{figure}[t!]
    \centering
  \includegraphics[width=\linewidth]{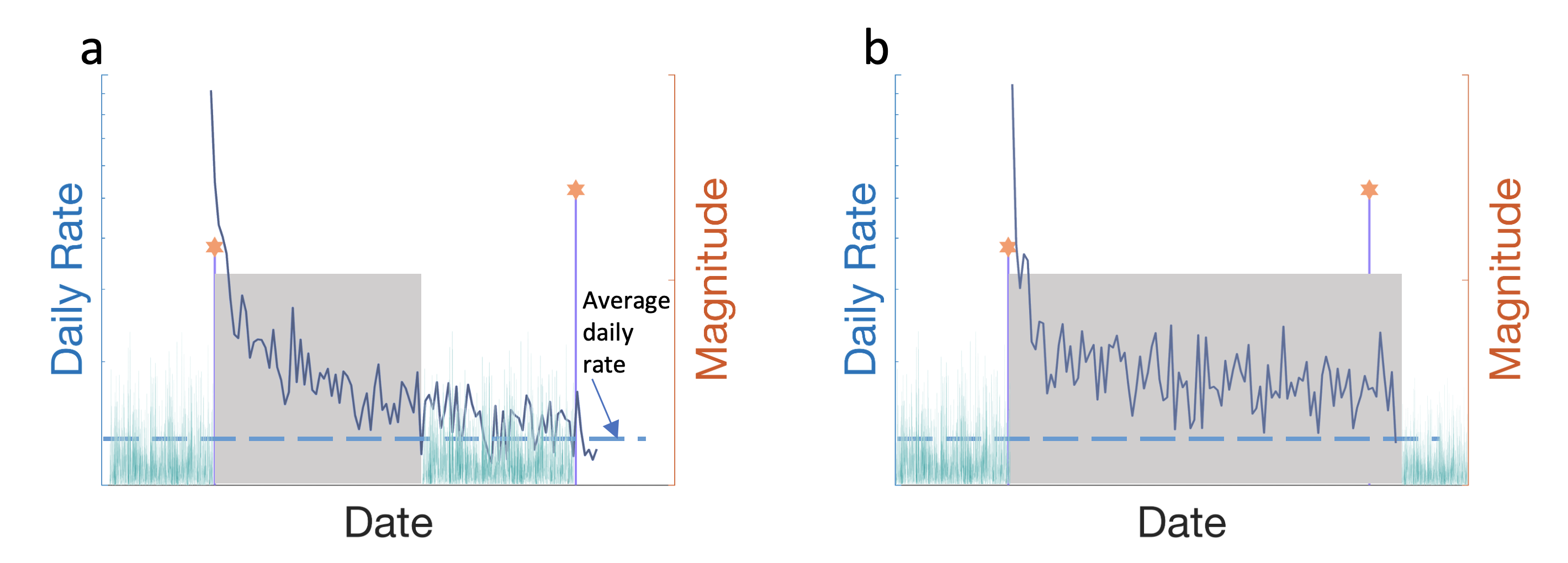}
    \caption{{\bf An illustrative diagram portrays two potential foreshock scenarios.} In both instances, a lesser-magnitude potential mainshock (depicted by the left orange Star of David) is followed by a more substantial potential mainshock (represented by the right orange Star of David). {\bf a}. The 10-day running-mean daily earthquake rate (blue curve) reverts to the daily mean earthquake rate (horizontal dashed light-blue line) before the occurrence of the larger mainshock; the aftershock window ends when the solid blue curve crosses the horizontal dashed light-blue line. Events within the aftershock period (cyan curve under gray shading) are excluded from the catalog, preserving the two mainshocks. {\bf b}. Similar to {\bf a}, but the aftershock period extends beyond the larger potential mainshock. Two cases are considered: I. The larger mainshock is omitted from the catalog as all earthquakes within the aftershock time window (gray shading, foreshock flag 1) are removed. II. The smaller mainshock is considered a foreshock, allowing earthquake events within the aftershock period (gray shading) to remain in the catalog (foreshock flag 0). Notably, scenarios akin to (b) are infrequent, and both foreshock flag approaches yield comparable outcomes.
}
    \label{fig:ff_cartoon}
\end{figure}

Then, we set another threshold, $M_{\rm th}$, typically $M_{\rm th}=4$ or $M_{th}=4.5$, and sort all the events with magnitudes greater than or equal to $M_{\rm th}$, to obtain a sorted list of events $(m_{p_1}\geq m_{p_2}\geq,..., \geq m_{p_r})$. Those events are now referred to as potential mainshocks, as some of them might fall within the aftershock period of a larger magnitude event. We consider the events starting with the highest magnitude and ending with the lowest. We proceed by computing the daily earthquake rate following the mainshock of interest. We ascertain the point at which this rate reverts to its baseline value (established as the average daily earthquake rate $\langle r_d \rangle$) using a 10-day moving window. We refer to the events during this time period when the rate is above the baseline as aftershocks and remove them. Sometimes mainshocks fall within the aftershock period and, thus, are deleted from the list of potential mainshocks.
This process of aftershocks removal is aimed to ensure that aftershocks from one event are not mistakenly interpreted as foreshocks of a subsequent event under examination; see Fig.~\ref{fig:ff_cartoon}.

In some cases, a mainshock $M_j$ falls in the aftershock window where its magnitude is larger than the magnitude of the mainshock under consideration $M_i$; i.e., $M_j>M_i$. In this case, there are two options: 1) The mainshock with the larger magnitude, $M_j$, is deleted as it can be considered as an aftershock; this case is referred to as a foreshock flag = 1. 2) Earthquake $i$ is considered a foreshock, and its corresponding aftershocks remain unaffected; this case is referred to as a foreshock flag = 0. There were only a few such cases, and the foreshock flag weakly affect the results described below. See the illustration in Fig. \ref{fig:ff_cartoon}.

After we repeat this process for all the possible mainshocks, we are left with a catalog like the one shown in Fig. \ref{fig:Fig1}(d), i.e., a set of identified mainshocks. We refer to the aftershock period of a mainshock $i$ as $D_{\mathrm{as}_i}$. We perform the analysis of the rate of earthquakes before mainshocks on the filtered catalog. In addition to the aforementioned approach, we also implemented a distance constraint of 100 kilometers from the epicenter region. The outcomes of this analysis were consistent with the results described in this here. See Fig. \ref{fig:SI9}.

\subsection*{Finding a significant number of events prior to mainshocks}

Once the mainshocks are identified, we proceed to count the number of events in the ``cleaned catalog'' that occurred before the mainshock, while taking into account the varying duration of the foreshock, denoted by $D_\mathrm{fs}$. Our null hypothesis is that the earthquake rate prior to a mainshock is not significantly different than the overall rate of the ``cleaned'' catalog---the null hypothesis is rejected when the actual rate is significantly different than the mean rate of the cleaned catalog. We show below that indeed the null hypothesis is rejected and the foreshock rate is significantly higher prior to mainshocks.

To estimate whether and by how much the rate prior to the mainshocks is larger than the mean rate, we calculate the rate prior to randomly selected times (during the time range spanned by the catalog). We then construct the probability density function (pdf) and the cumulative density function (cdf) of these rates. Then, we calculate how many events fall above a certain percentile (for example, 95\%) of the surrogate data. If the ratio of recorded events above the specific percentile is higher than the expected one (based on the surrogate data), we can conclude that the null hypothesis is rejected and that the rate of events preceding the mainshock is indeed larger than expected; this can have predictive power. We repeated this procedure using different percentiles, such as 50\% (median), 80\%, 95\%, and 99\%. See Fig. \ref{fig:Fig2} for an example of the results for a time window of $D_{\mathrm{fs}}=3$ days.

\subsection{Earthquake catalogs}

In this study, we analyze two catalogs, the Hauksson { et al.} catalog \cite{hauksson2012waveform,li2018machine} and the Ross { et al.} \cite{ross2019searching} catalog. The Hauksson { et al.} catalog can be obtained from the Southern California Earthquake Data Center (see the Availability of data in the materials section). It spans 40 years, 1981-2019, and is considered to be complete for magnitudes $m\geq2.5$. It contains $N = 45843$ events with magnitudes larger than or equal to the magnitude threshold of 2.5. The catalog covers the area of Southern California, between 113.77\degree{W}-122.16{W} and 30.16\degree{N}--37.51\degree{N}. The Ross { et al.} catalog spans a 10-year period from 01-Jan-2008 to 31-Dec-2017 and is nearly complete for magnitudes above 0.3 \cite[][]{ross2019searching}. To ensure completeness, we chose the magnitude threshold to be $m_c=1$ and excluded all events below this magnitude, ending up with $N = 171,296$ events. The catalog covers the area of Southern California, between 31.9\degree{N}--37.03\degree{N} and 114.81\degree{W}--121.71\degree{W}.

\subsection{The ETAS Model}
The ETAS model \cite{ogata1988statistical} describes earthquake aftershocks based on the analogy of an epidemic. The “infected” individual (or the immigrant) enters the observation region and spreads the infection (or the immigrant produces offspring). Immigrants are analogous to background events, and their offspring to aftershocks. In this sense, the model can also be thought of as an ancestor-offspring system. Likewise, these offspring can reproduce (i.e., spread infection). After most of the population has been infected or died, there are insufficient receptive individuals for the epidemic to continue. According to the seismological analogy, this would be equivalent to an aftershock sequence (such as an immigrant family) initiated after a background event (such as an ancestor) had died out. As of now, the ETAS model is one of the most common operational earthquake models.

The ETAS model assumes that seismic events involve a stochastic point process in space-time \cite[][]{ogata1988statistical,ogata1998space}. Each event above a magnitude $M_0$ is selected independently from the Gutenberg–Richter distribution (where b = 1). The conditional rate, $\lambda$, at location $(x, y)$ at time $t$ is given by
\begin{equation}
    \lambda(x,y,t | H_t) = \mu(x,y) + \sum_{t<t_i} k(M_i)\exp(\alpha(M_i-M_0))/(t - t_i + c)^p
\end{equation}
where $c$ and $p$ are parameters in the Omori-Utsu law, $k$ and $\alpha$ control the aftershock productivity by a mainshock and its magnitude sensitivity, respectively. These five parameters, $\mu, c, p, k$ and $\alpha$, can be determined using the Maximum Likelihood Estimation (MLE) with an earthquake catalog consisting of hypo-central times $t_i$ and magnitudes $M_i$ which is denoted by $H_t$.
The estimated parameters are given in Table \ref{Tab:ETAS_params}.

\section{Results}
\textbf{Number of events prior to a mainshock compared with surrogate catalogs.}
We systematically study the number of recorded events $n$ (foreshocks) prior to identified mainshocks using different time windows of $D_{\mathrm{fs}} = 1,2,3,5,10,20,30,40$ days. For each mainshock $i$, let $n_i$ be the number of foreshocks that occurred during the time duration $D_{\mathrm{fs}_i}$ prior to mainshock $i$.
Using the surrogate test described above, we compare the distributions of the real and the surrogate counts and test whether the counts before the mainshocks are larger than the corresponding counts of the control surrogate data.

In addition to the surrogate data test, we use the Poisson distribution to estimate the significance of the foreshock counts in the cleaned catalog; a Poisson process is often assumed to model the earthquake background rate. A significant number of events above the level of the expected percentile indicates that the null hypothesis that the earthquake rate prior to the mainshock is similar to the mean catalog rate is rejected. This suggests an increased seismic activity prior to strong events, in comparison to an uncorrelated Poisson process. The Poisson distribution is given by $f(k;\lambda) = \frac{\lambda^k e^{-\lambda}}{k!}$ where $\lambda$ is the mean of the entire (cleaned) catalog.

\begin{figure}[ht!]
\centering
	\includegraphics[width=\linewidth]{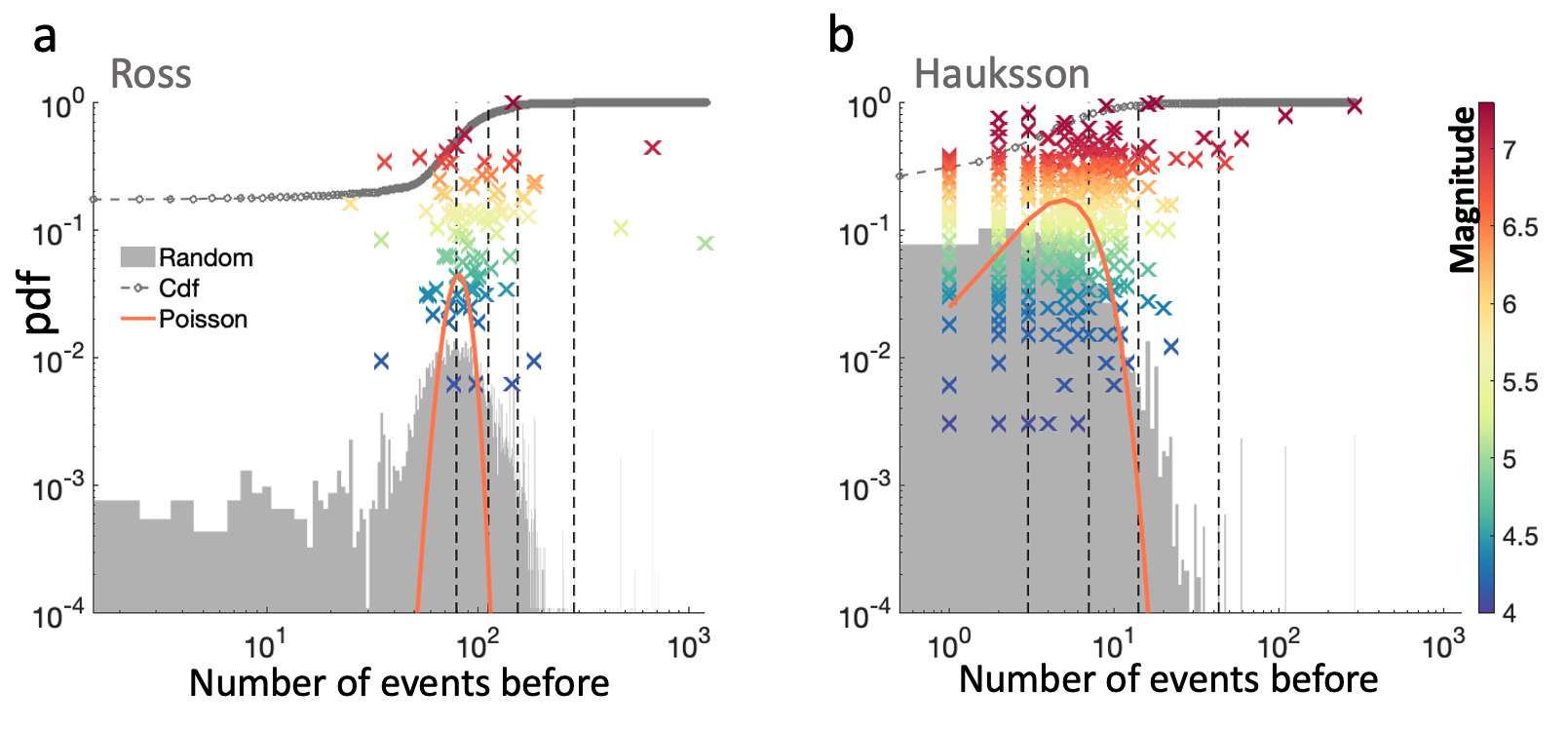}
	\caption{ {\bf The number of events prior to mainshocks.} The number of events recorded before a mainshock (marked by colored crosses), in a period of $D_{\mathrm{fs}} = 3$ days; the colors represent the magnitude of the mainshocks. The grey bars represent the pdf of the number of events recorded within 1000 randomly selected time windows of three days; this serves as a surrogate test. We plot the corresponding cumulative distribution function (cdf) in the gray circles. The dashed vertical black lines represent the 50\%, 80\%, 95\%, and 99\% quantiles of the surrogate distribution. \textbf{a}. Results of analysis of the Ross { et al.} catalog, when the magnitude threshold is 1. The mean number of foreshocks for the real (surrogate) mainshocks is 120 (81) events. Here, 3 out of the 92 detected mainshocks fall above the $99\%$ quantile, suggesting more than a threefold increase from the expected 1\%. \textbf{b}. Same as {\bf a}, for the Hauksson { et al.} catalog whose magnitude threshold is 2.5. The number of detected mainshocks is 421. Here, the mean number of foreshocks for the real (surrogate) mainshocks is 7 (5.3) events. Only four events fall above the 99\% quantile, which is smaller than the expected 1\%. The orange line represents the Poisson distribution, which is calculated based on the mean rate of the clean catalogs. }
	\label{fig:Fig2}
\end{figure}

In Fig. \ref{fig:Fig2}(a), we show that for fixed time window $D_{\mathrm{fs}}=3$ days, most of the mainshocks are preceded by more than 100 events in 3 days, with an average of $166$ events. This number is significantly higher than the number we obtained for the surrogate events (mean value of 102 events); i.e., $\approx$60\% increase. 
The results for $D_{\mathrm{fs}}=1,2,5,10$ days yielded qualitatively similar results. To evaluate the significance of the results, we calculated the $50\%, 80\%,95\%$, and $99\%$ quantiles of the surrogate results distribution.
In Fig. \ref{fig:Fig2}, we present the summary of results for the Ross { et al.} and Hauksson { et al.} catalogs, respectively, where the dashed vertical lines represent the different quantiles, and the colored ``x'' represents the number of events prior to the different mainshocks. The pdf of the surrogate data is marked by the gray bars, and the gray circles mark the cdf of the surrogate data. 


After calculating the quantiles, we examined how many events fall above each quantile. For example, when considering mainshocks with a magnitude larger than or equal to 4, using the Ross { et al.} catalog, 89 events out of 92 potential mainshocks were identified as mainshocks (while the other three are now considered as aftershocks). For the $99\%$ quantile of the surrogate data, we counted three events with more foreshocks than the $99\%$ of the random events; i.e., 3/89=3.4\% real mainshocks, over three times the expected 1\% based on the surrogate data. For the 95\% quantile, 10 out of 89 detected mainshocks fell above or equal to the quantile, i.e., 10/89 = 11\%, which is more than twice as many events as expected $5\%$.

\begin{figure}[ht!]
\centering
	\includegraphics[width=0.9\linewidth]{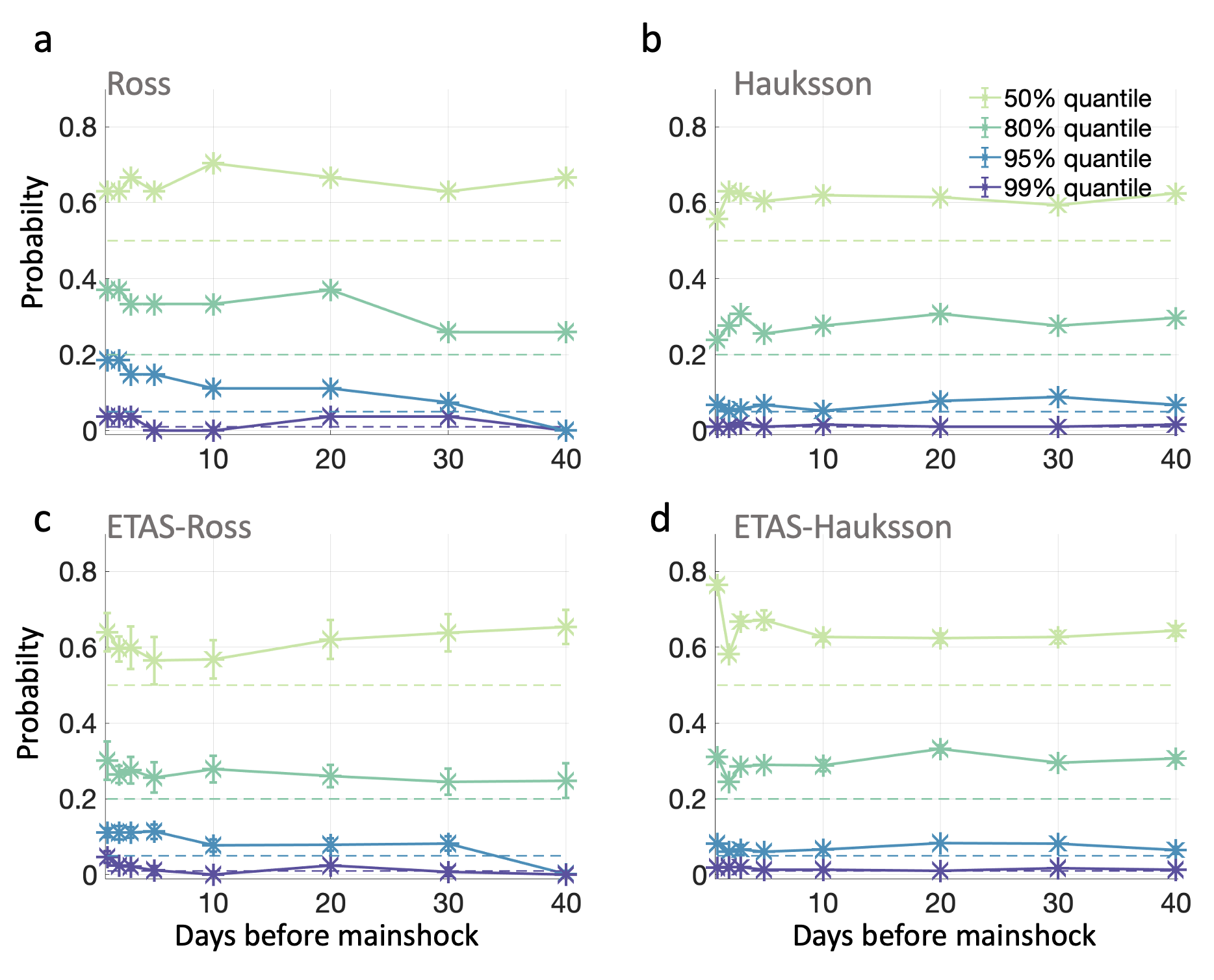}
	\caption{ {\bf The probability of the number of foreshocks occurring above a certain quantile of the surrogate data pdf for different time windows.} The different colors represent the results of the different quantiles as indicated in the legend.
The solid lines with stars depict the results that are based on the data with respect to the surrogate data. The dashed horizontal colored lines indicate the 1\%, 5\%, 20\%, and 50\% levels (corresponding to the 99\%, 95\%, 80\%, and 50\% quantiles) such that the symbols that appear above their corresponding dashed lines indicate a higher rate than expected. Results of the {\bf a}. Ross { et al.} and {\bf b}. Hauksson { et al.} catalogs, when considering mainshocks with magnitudes larger than or equal to 4.5, and when the magnitude threshold is 1 and 2.5 for the Ross { et al.} and Hauksson { et al.} catalogs, respectively. The foreshock flag is 0 (i.e., a strong foreshock does not delete a mainshock). Here, for example, for the Ross { et al.} catalog, when the time window is two days, 19\% of the events fall above the $95\%$ quantile, which is almost a four-fold increase over the expected number of foreshocks. Note that the results for the Ross { et al.} catalog are higher than the results for the Hauksson { et al.} catalog. The results (mean $\pm$ 1 std) were obtained based on 10 synthetic ETAS model catalogs, with parameters estimated using the \textbf{c}. Ross { et al.} and \textbf{d}. Hauksson { et al.} catalogs. The ETAS model that is based on the Ross { et al.} catalog results shows smaller errors, indicating greater confidence. 
It should be noted that for certain instances, the error bars presented in panels c and d may not be discernible due to their small size.}
\label{fig:Fig3}
\end{figure}

We summarize the results of the Ross { et al.} catalog in Fig.~\ref{fig:Fig3}(a). It is clear that for a time window of fewer than five days, there are 2--3 times more events prior to the mainshocks in comparison to the surrogate data results, for the higher quantiles. This indicates that the null hypothesis that the earthquake rate prior to mainshocks is similar to the mean rate is rejected and that earthquake activity increases prior to mainshocks. 
 
We also analyzed the Hauksson { et al.} catalog whose magnitude threshold is 2.5.
Here (Fig. \ref{fig:Fig3}b) the percentages of events above or equal to the $95\%$ and $80\%$ quantiles are 0.06 and 0.25, respectively, much lower than the fraction obtained for the Ross { et al.} catalog (Fig. \ref{fig:Fig3}a). We attribute the more significant results of the Ross { et al.} catalog to the lower magnitude threshold that provides many more events, thus enabling easy detection of the increased earthquake rate prior to mainshocks. 

\textbf{Comparison against the ETAS model.} We generated 10 synthetic ETAS model catalogs, using the parameters fitted for the Hauksson { et al.} and Ross { et al.} catalogs, and using the MLE \cite[][]{ogata1988statistical}; see Table \ref{Tab:ETAS_params} and \url{https://www.kaggle.com/code/tokutani/etas-model-analysis-for-japan-region}. Next, we tested our method on these synthetic catalogs, and the results are summarized in Fig. \ref{fig:Fig3}c,d. In general, the probabilities in the ETAS model are smaller than those of the real data, indicating that some earthquake processes are missing in the ETAS model and that the current ETAS model formulation cannot fully reproduce the increased activity prior to mainshocks. We note that the results of the 99\% quantile of the real data may be more uncertain due to the few events that fell above this quantile. We performed the same analysis when foreshock flag=1 and obtained similar results; see Fig.~\ref{fig:SI_Fig1_Omori}. This indicates that the strict definition of foreshocks within the set of potential mainshocks is not important.

\section{Discussion}
In our study, we focus on the inverse Omori law and show that indeed, statistically, the earthquake rate prior to mainshocks, is significantly larger than usual. Furthermore, we analyze the conventional Omori-Utsu law and summarize the results in the Supplementary Material. We show that the Omori-Utsu law does not hold for the vast majority of mainshocks. However, an analysis of ETAS model synthetic catalogs yields results similar to those for the real catalogs. Since the ETAS model is based on the Omori-Utsu law and since this law is valid only for a relatively small portion of the events, we conclude that the Omori-Utsu law is valid in a statistical sense and probably underlies the rate of aftershocks after mainshocks. 

In summary, we mainly examine the inverse and, to a lesser degree, the conventional versions of the Omori-Utsu law, using a highly detailed catalog. We also develop and implement an algorithm for mainshock detection. The algorithm is composed of sorting the earthquakes according to their magnitude and deleting the aftershocks appearing between the mainshock and the time point at which the earthquake rate returns to the mean catalog rate. This algorithm also allows us to define the foreshocks.

We then propose a surrogate test to investigate whether mainshocks are often preceded by more events than occur during random time windows. This is done by comparing the earthquake rate within a certain time window prior to a mainshock to the earthquake rate of randomly selected time windows over the entire ``cleaned'' catalog. By analyzing two catalogs that differ in their magnitude completeness, we show that the more detailed catalog resulted in a greater probability for higher earthquake rates prior to mainshocks, sometime three times higher than expected. [Note that it is possible, in principle, to analyze the {Ross et al.} catalog, instead of the {Hauksson et al.} catalog, when setting the magnitude threshold to 2.5 instead of 1. However, this will result in a catalog with a small number of events as this catalog spans 10 years, while the catalog of {Hauksson et al.} spans 40 years of events, thus allowing a much better statistical analysis. This is why we use this catalog when dealing with a higher magnitude threshold.] In principle, the increased earthquake rate prior to mainshocks can be used to provide some forecasting of a mainshock. For instance, if the probability to be above the 99\% percentile of the surrogate data is 3\% (instead of 1\%), then an increased rate within the upper 1\% indicates, with high probability (three times more than expected), indicate a near future major earthquake, within the time window of the analysis.

 Recently, Trugman and Ross \cite{trugman2019pervasive} analyzed the detailed catalog of Ross et al. \cite{ross2019searching} and found an increased earthquake rate prior to mainshocks. Our research not only supports the conclusions drawn by \cite{trugman2019pervasive} but also contributes several new insights: i) Instead of presuming the null hypothesis of gamma-distributed interevent intervals, we rely on the empirical data.
ii) Our analysis is not confined to a 10 km radius from the epicenter, demonstrating that the increased seismicity rate before mainshocks can be detected across the entire southern California region.
iii) We establish that the Epidemic-Type Aftershock Sequence (ETAS) model fails to replicate the heightened earthquake rate observed prior to mainshocks.
iv) We demonstrate that the elevated rate is not discernible when employing the less comprehensive catalog by Hauksson et al. \cite{hauksson2012waveform}. And 
v) we validate the reported findings using our alternative methodology for identifying foreshocks, mainshocks, and aftershocks.

An increased foreshock rate before a large seismic event may be associated with the accumulation of damage (stress) before failure (a quake). This is based on both theoretical \cite[][]{lyakhovsky2008scaling} and lab experiments \cite[][]{lockner1993role}. Foreshock activity may vary greatly from fault to fault depending on several factors, such as fault maturity and fault heat flux \cite[][]{ben2003characterization}, reflecting the damage accumulation in each fault system prior to the mainshock. This is supported by theoretical studies that predict that there should be some degree of damage before the fault is activated \cite[][]{ben2006analysis}.

In addition, we tested and compared synthetic catalogs generated using the ETAS model. We show that this model does not fully reproduce the results obtained for real catalogs, suggesting that some key processes regarding the development of an earthquake are missing. Finally, we tested the conventional version of the Omori-Utsu law for aftershocks, which describes the rate of aftershocks after mainshocks. Our results suggest that this law appears to hold only for a few mainshocks, suggesting that models should not heavily rely on this law.

\subsection*{Acknowledgments}
We thank Warner Marzocchi for fruitful discussions.

\subsection*{Funding}
We thank the EU H2020 project RISE, the Israel Science Foundation (grant no. 189/19), the Israel Ministry of Energy, and the PAZY Foundation for financial support.
    
\subsection*{Availability of data and materials}
Data of the Ross { et al.} catalog can be downloaded at \url{ https://www.science.org/doi/10.1126/science.aaw6888}. 
Data of Hauksson {\it et al.} catalog can be downloaded at \\
\url{ https://scedc.caltech.edu/data/alt-2011-dd-hauksson-yang-Hauksson.html}. 
The code for main methods used in this project is available at \url{https://github.com/eitanas/Omori-Utsu-Law-Revisited}.

\subsection*{Authors' contributions}
Y.A. conceptualized, developed, and supervised the project. S.H supervised the project. S.H and S.M. gave useful suggestions. E.E.A. implemented the methodology, performed the analysis, and prepared the figures. E.E.A. and Y.A. wrote the manuscript.

\bibliographystyle{unsrt}
\bibliography{INCREASED_EQ_RATE_PRIOR_TO_MAINSHOCKS.bib}

\begin{thebibliography}{10}

\bibitem{jordan2006earthquake}
Thomas~H Jordan.
\newblock Earthquake predictability, brick by brick.
\newblock {\em Seismological Research Letters}, 77(1):3--6, 2006.

\bibitem{jordan2011operational}
T~Jordan, Y-T Chen, Paolo Gasparini, Raul Madariaga, Ian Main, Warner
  Marzocchi, Gerassimos Papadopoulos, K~Yamaoka, J~Zschau, et~al.
\newblock Operational earthquake forecasting: State of knowledge and guidelines
  for implementation.
\newblock {\em Annals of Geophysics}, 2011.

\bibitem{ogata2017statistics}
Yosihiko Ogata.
\newblock Statistics of earthquake activity: Models and methods for earthquake
  predictability studies.
\newblock {\em Annual Review of Earth and Planetary Sciences}, 45:497--527,
  2017.

\bibitem{ogata1988statistical}
Yosihiko Ogata.
\newblock Statistical models for earthquake occurrences and residual analysis
  for point processes.
\newblock {\em Journal of the American Statistical association}, 83(401):9--27,
  1988.

\bibitem{zhang2021improved}
Yongwen Zhang, Dong Zhou, Jingfang Fan, Warner Marzocchi, Yosef Ashkenazy, and
  Shlomo Havlin.
\newblock Improved earthquake aftershocks forecasting model based on long-term
  memory.
\newblock {\em New Journal of Physics}, 23(4):042001, 2021.

\bibitem{gutenberg1944frequency}
Beno Gutenberg and Charles~F Richter.
\newblock Frequency of earthquakes in california.
\newblock {\em Bulletin of the Seismological society of America},
  34(4):185--188, 1944.

\bibitem{gutenberg1956earthquake}
Beno Gutenberg and Carl~F Richter.
\newblock Earthquake magnitude, intensity, energy, and acceleration: (second
  paper).
\newblock {\em Bulletin of the seismological society of America},
  46(2):105--145, 1956.

\bibitem{omori1894after}
Fusakichi Omori.
\newblock On the after-shocks of earthquakes.
\newblock {\em J. Coll. Sci., Imp. Univ., Japan}, 7:111--200, 1894.

\bibitem{utsu1961statistical}
Tokuji Utsu.
\newblock A statistical study on the occurrence of aftershocks.
\newblock {\em Geophys. Mag.}, 30:521--605, 1961.

\bibitem{ogata1998space}
Yosihiko Ogata.
\newblock Space-time point-process models for earthquake occurrences.
\newblock {\em Annals of the Institute of Statistical Mathematics},
  50(2):379--402, 1998.

\bibitem{gerstenberger2005real}
Matthew~C Gerstenberger, Stefan Wiemer, Lucile~M Jones, and Paul~A Reasenberg.
\newblock Real-time forecasts of tomorrow's earthquakes in california.
\newblock {\em Nature}, 435(7040):328--331, 2005.

\bibitem{woessner2011retrospective}
J~Woessner, Sebastian Hainzl, W~Marzocchi, MJ~Werner, AM~Lombardi, F~Catalli,
  B~Enescu, M~Cocco, MC~Gerstenberger, and S~Wiemer.
\newblock A retrospective comparative forecast test on the 1992 landers
  sequence.
\newblock {\em Journal of Geophysical Research: Solid Earth}, 116(B5), 2011.

\bibitem{taroni2018prospective}
Matteo Taroni, Warner Marzocchi, Danijel Schorlemmer, Maximilian~Jonas Werner,
  Stefan Wiemer, Jeremy~Douglas Zechar, Lukas Heiniger, and Fabian Euchner.
\newblock Prospective csep evaluation of 1-day, 3-month, and 5-yr earthquake
  forecasts for italy.
\newblock {\em Seismological Research Letters}, 89(4):1251--1261, 2018.

\bibitem{shcherbakov2004generalized}
Robert Shcherbakov, Donald~L Turcotte, and John~B Rundle.
\newblock A generalized omori's law for earthquake aftershock decay.
\newblock {\em Geophysical research letters}, 31(11), 2004.

\bibitem{davidsen2015generalized}
J{\"o}rn Davidsen, Chad Gu, and Marco Baiesi.
\newblock Generalized omori--utsu law for aftershock sequences in southern
  california.
\newblock {\em Geophysical Journal International}, 201(2):965--978, 2015.

\bibitem{papazachos1973time}
BC~Papazachos.
\newblock The time distribution of the reservoir-associated foreshocks and its
  importance to the prediction of the principal shock.
\newblock {\em Bulletin of the Seismological Society of America},
  63(6-1):1973--1978, 1973.

\bibitem{kagan1978statistical}
Yan Kagan and Leon Knopoff.
\newblock Statistical study of the occurrence of shallow earthquakes.
\newblock {\em Geophysical Journal International}, 55(1):67--86, 1978.

\bibitem{jones1979some}
Lucile~M Jones and Peter Molnar.
\newblock Some characteristics of foreshocks and their possible relationship to
  earthquake prediction and premonitory slip on faults.
\newblock {\em Journal of Geophysical Research: Solid Earth},
  84(B7):3596--3608, 1979.

\bibitem{console1993foreshock}
Rodolfo Console, Maura Murru, and Bruno Alessandrini.
\newblock Foreshock statistics and their possible relationship to earthquake
  prediction in the italian region.
\newblock {\em Bulletin of the Seismological Society of America},
  83(4):1248--1263, 1993.

\bibitem{chen2016analysis}
Xiaowei Chen and Peter~M Shearer.
\newblock Analysis of foreshock sequences in california and implications for
  earthquake triggering.
\newblock {\em Pure and Applied Geophysics}, 173(1):133--152, 2016.

\bibitem{moutote2021rare}
Luc Moutote, David Marsan, Olivier Lenglin{\'e}, and Zacharie Duputel.
\newblock Rare occurrences of non-cascading foreshock activity in southern
  california.
\newblock {\em Geophysical research letters}, 48(7):e2020GL091757, 2021.

\bibitem{trugman2019pervasive}
Daniel~T Trugman and Zachary~E Ross.
\newblock Pervasive foreshock activity across southern california.
\newblock {\em Geophysical Research Letters}, 46(15):8772--8781, 2019.

\bibitem{ross2019searching}
Zachary~E Ross, Daniel~T Trugman, Egill Hauksson, and Peter~M Shearer.
\newblock Searching for hidden earthquakes in southern california.
\newblock {\em Science}, 364(6442):767--771, 2019.

\bibitem{manganiello2021anomalous}
Ester Manganiello, Marcus Herrmann, and Warner Marzocchi.
\newblock Anomalous foreshock activity in southern {California} is associated
  with zones of high heat flow.
\newblock {\em Earth and Space Science Open Archive}, page~13, 2021.

\bibitem{xie2020promise}
Yazhou Xie, Majid Ebad~Sichani, Jamie~E Padgett, and Reginald DesRoches.
\newblock The promise of implementing machine learning in earthquake
  engineering: A state-of-the-art review.
\newblock {\em Earthquake Spectra}, 36(4):1769--1801, 2020.

\bibitem{li2018machine}
Zefeng Li, Men-Andrin Meier, Egill Hauksson, Zhongwen Zhan, and Jennifer
  Andrews.
\newblock Machine learning seismic wave discrimination: Application to
  earthquake early warning.
\newblock {\em Geophysical Research Letters}, 45(10):4773--4779, 2018.

\bibitem{beroza2021machine}
Gregory~C Beroza, Margarita Segou, and S~Mostafa~Mousavi.
\newblock Machine learning and earthquake forecasting—next steps.
\newblock {\em Nature communications}, 12(1):1--3, 2021.

\bibitem{tan2021machine}
Yen~Joe Tan, Felix Waldhauser, William~L Ellsworth, Miao Zhang, Weiqiang Zhu,
  Maddalena Michele, Lauro Chiaraluce, Gregory~C Beroza, and Margarita Segou.
\newblock Machine-learning-based high-resolution earthquake catalog reveals how
  complex fault structures were activated during the 2016--2017 central italy
  sequence.
\newblock {\em The Seismic Record}, 1(1):11--19, 2021.

\bibitem{hauksson2012waveform}
Egill Hauksson, Wenzheng Yang, and Peter~M Shearer.
\newblock Waveform relocated earthquake catalog for southern california (1981
  to june 2011).
\newblock {\em Bulletin of the Seismological Society of America},
  102(5):2239--2244, 2012.

\bibitem{lyakhovsky2008scaling}
Vladimir Lyakhovsky and Yehuda Ben-Zion.
\newblock Scaling relationjones1979somes of earthquakes and aseismic
  deformation in a damage rheology model.
\newblock {\em Geophysical Journal International}, 172(2):651--662, 2008.

\bibitem{lockner1993role}
December Lockner.
\newblock The role of acoustic emission in the study of rock fracture.
\newblock In {\em International Journal of Rock Mechanics and Mining Sciences
  \& Geomechanics Abstracts}, volume~30, pages 883--899. Elsevier, 1993.

\bibitem{ben2003characterization}
Yehuda Ben-Zion and Charles~G Sammis.
\newblock Characterization of fault zones.
\newblock {\em Pure and applied geophysics}, 160(3):677--715, 2003.

\bibitem{ben2006analysis}
Yehuda Ben-Zion and Vladimir Lyakhovsky.
\newblock Analysis of aftershocks in a lithospheric model with seismogenic zone
  governed by damage rheology.
\newblock {\em Geophysical Journal International}, 165(1):197--210, 2006.

\end{thebibliography}

\appendix
\clearpage\newpage
\section*{Supplemental Information}
\setcounter{figure}{0}
\renewcommand{\thefigure}{S\arabic{figure}}

\begin{center}
\begin{table}[ht!]
\begin{tabular}{|c|c|c|c|c|c|}
\hline
Parameter & $\mu$ & A & c & $\alpha$ & p  \\
\hline
Ross & 0.96 & 50 & 0.0001 & 1.98 & 1.29 \\
Hauksson & 0.58 & 2.5  & 0.01 & 1.83 & 1.18 \\
\hline
\end{tabular} 
\caption{{\bf ETAS model estimated parameters.} The parameters were evaluated using the MLE applied to the Hauksson et al. and Ross et al. catalogs, using \protect\url{https://www.kaggle.com/code/tokutani/etas-model-analysis-for-japan-region}.}
\label{Tab:ETAS_params}
\end{table}
\end{center}


\section{Omori-Utsu law.}
\begin{figure}[ht!]
\centering
	\includegraphics[width=\linewidth ]{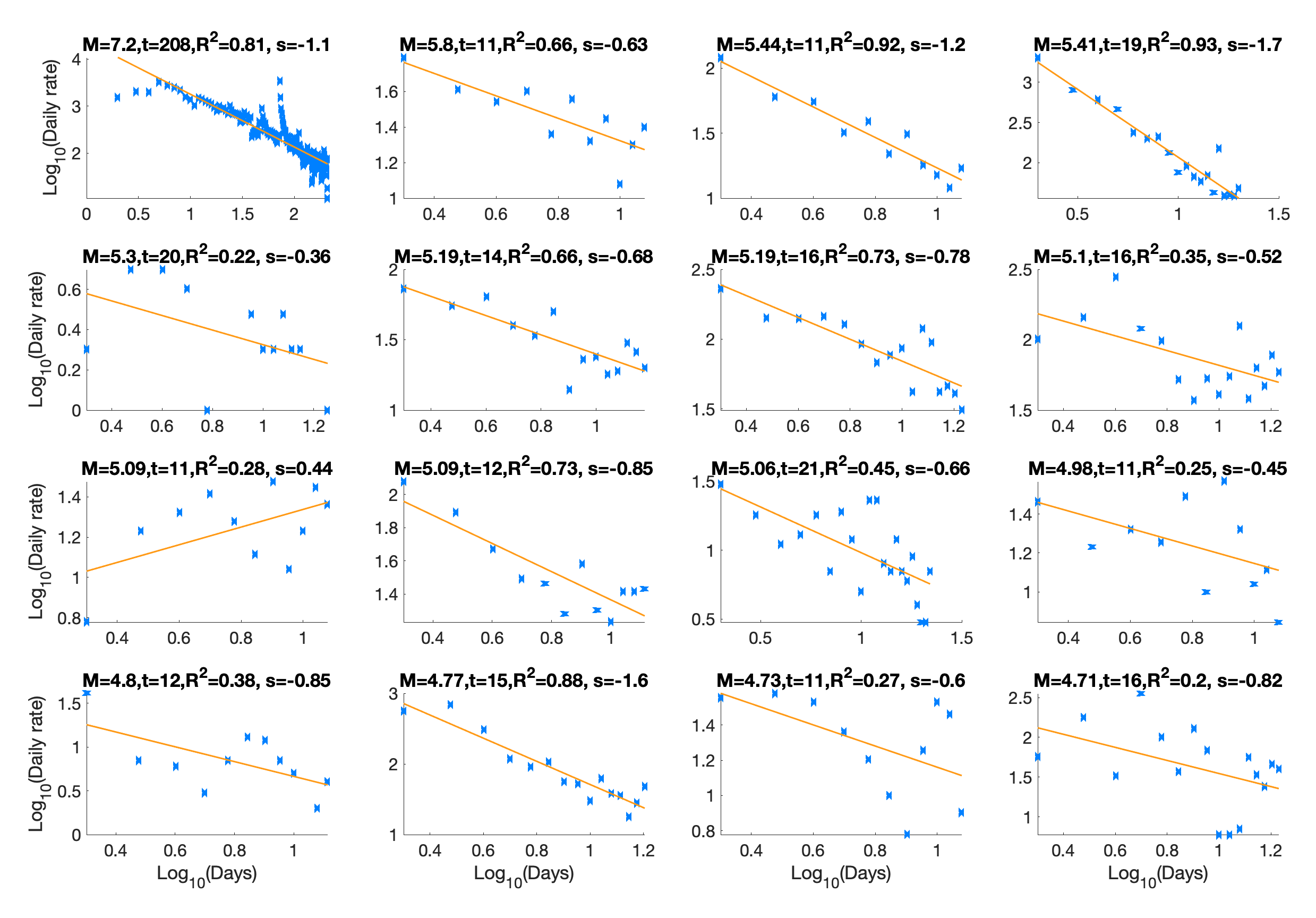}
	\caption{
        {\bf The Omori-Utsu law.} Daily rate as a function of days after a mainshock in a log-log scale is presented for the identified mainshocks. We performed a linear regression (on a log-log plot) and calculated the curve slope (exponent) and its goodness of fit, $R^2$. We show only 16 mainshock curves with $R^2>0.1$.
	On the title of each subplot, we add the following information: `M'--the events' magnitude, `t'--the events' relaxation time, $R^2$, and `s'--the slope of the linear fit (or, exponent). 
	}
	\label{fig:SI_Fig1_Omori_Real}
\end{figure}

In this part, we examine the validity of the conventional Omori-Utsu law, which states that the earthquake rate after mainshocks decreases as a power law with an exponent of about -1. Following the detection of each mainshock, we calculate the event rate decrease in this period. We estimate the power law exponent of the decay rate using a linear fit on a log-log plot. We also calculate the goodness of fit $R^2$. The results of 16 mainshocks for which the slope is relatively significant (i.e., $R^2>0.1$) are presented in Fig. \ref{fig:SI_Fig1_Omori_Real}; only for a few mainshocks is the power law exponent close to -1, and in one case, the slope is positive instead of negative. 

\begin{figure}[ht!]
\centering
\includegraphics[width=\linewidth]{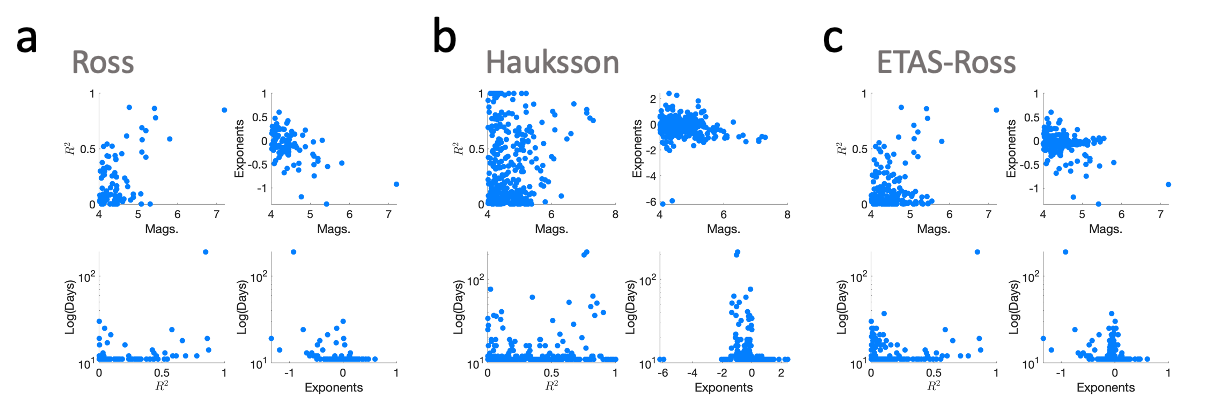}
	\caption{
	{\bf Aftershock decay rate exponent, $R^2$, and relaxation time.}
	We summarize the results for all the mainshocks that were detected in the \textbf{a}. Ross { et al.} catalog, \textbf{b}. Hauksson {et al.} catalog, and \textbf{c}. the ETAS model synthetic catalogs (based on Ross parameters). The number of mainshocks that were detected in the Ross { et al.} and the Hauksson { et al.} catalogs are 89 and 394, respectively.
	The four subplots of the catalogs describe (clockwise, starting at the top left) the following: magnitude of the event plotted vs. the $R^2$,  magnitude of the event vs. the decay exponent, relaxation time (in days) vs. $R^2$, and relaxation time vs. the exponent. Clearly, it is difficult to validate the Omori-Utsu law from the results.
	}
\label{fig:SI_Fig1_Omori_Real_summary}
\end{figure}

We summarize the results of all mainshocks in \ref{fig:SI_Fig1_Omori_Real_summary} where we show the exponent and $R^2$ for both the Ross {et al.} and Hauksson {et al.} catalogs. It is apparent that the slopes are both negative and positive and that there is no clear power law decay of the earthquake rate after mainshocks. Moreover, there is no clear correlation between the magnitude of the mainshocks and the rate decaying exponent and between the decay time and the exponent/slope. We thus conclude that the Omori-Utsu law is clearly valid in only a small portion of the identified mainshocks. We repeat this analysis on synthetic catalogs generated by the ETAS model, and the results are presented in Fig. \ref{fig:SI_Fig1_Omori_Real_summary}. It is clear that the results that are based on the synthetic ETAS models catalogs are similar to the results that are based on the real catalog. The ETAS model is constructed using known laws, including the Omori-Utsu law. As such, one would expect to find in the synthetic ETAS model catalogs, many mainshocks that follow clearly the Omori-Utsu law. The difficulty in identifying this law in many mainshocks might be related to independent mainshocks that interfere and break the aftershock sequence of other mainshocks.It is possible to achieve greater validity of the Omori-Utsu law by focusing on earthquakes that occur near the epicenters of the mainshocks. However, this appears to be an unlikely outcome. The above suggests that the Omori-Utsu law underlies the aftershock activity, in both real and synthetic catalogs, despite the fact that it is not visible for most mainshocks.   


\begin{figure}[ht!]
\centering	\includegraphics[width=\linewidth]{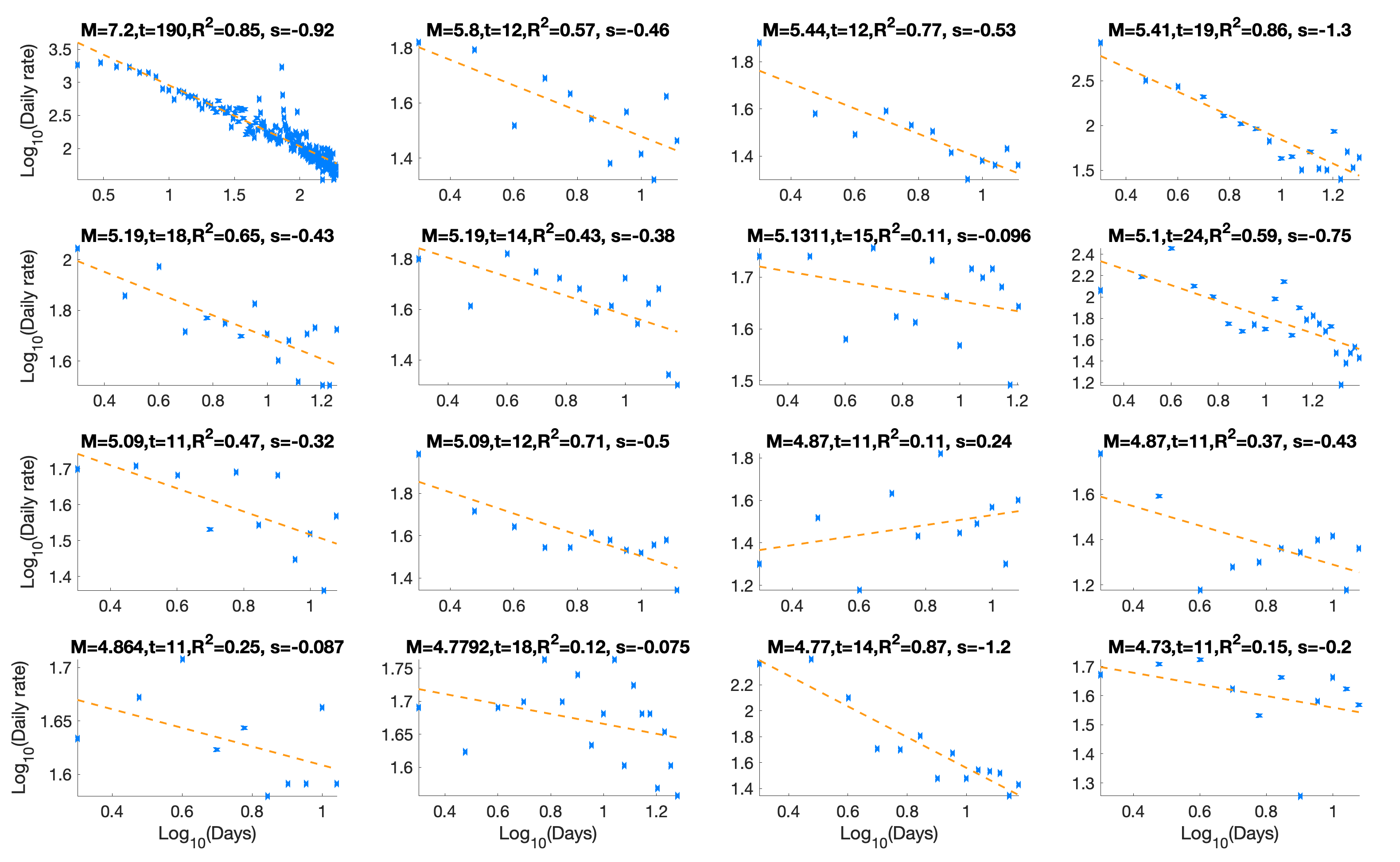}
	\caption{
        {\bf The Omori-Utsu law tested on ETAS synthetic catalogs.} Daily rate as a function of days after a mainshock in a log-log scale is presented for the identified mainshocks. We performed a linear regression (on a log-log plot) and calculated the curve slope (exponent) and its goodness of fit, $R^2$. We show only 16 mainshock curves with $R^2>0.1$.
	On the title of each subplot, we add the following information: `M'--the events' magnitude, `t'--the events' relaxation time, $R^2$, and `s'--the slope of the linear fit (or, exponent). 
	}
	\label{fig:SI_Fig3_Omori_ETAS}
\end{figure}

\newpage


\begin{figure}[ht!]
\centering
	\includegraphics[width=\linewidth]{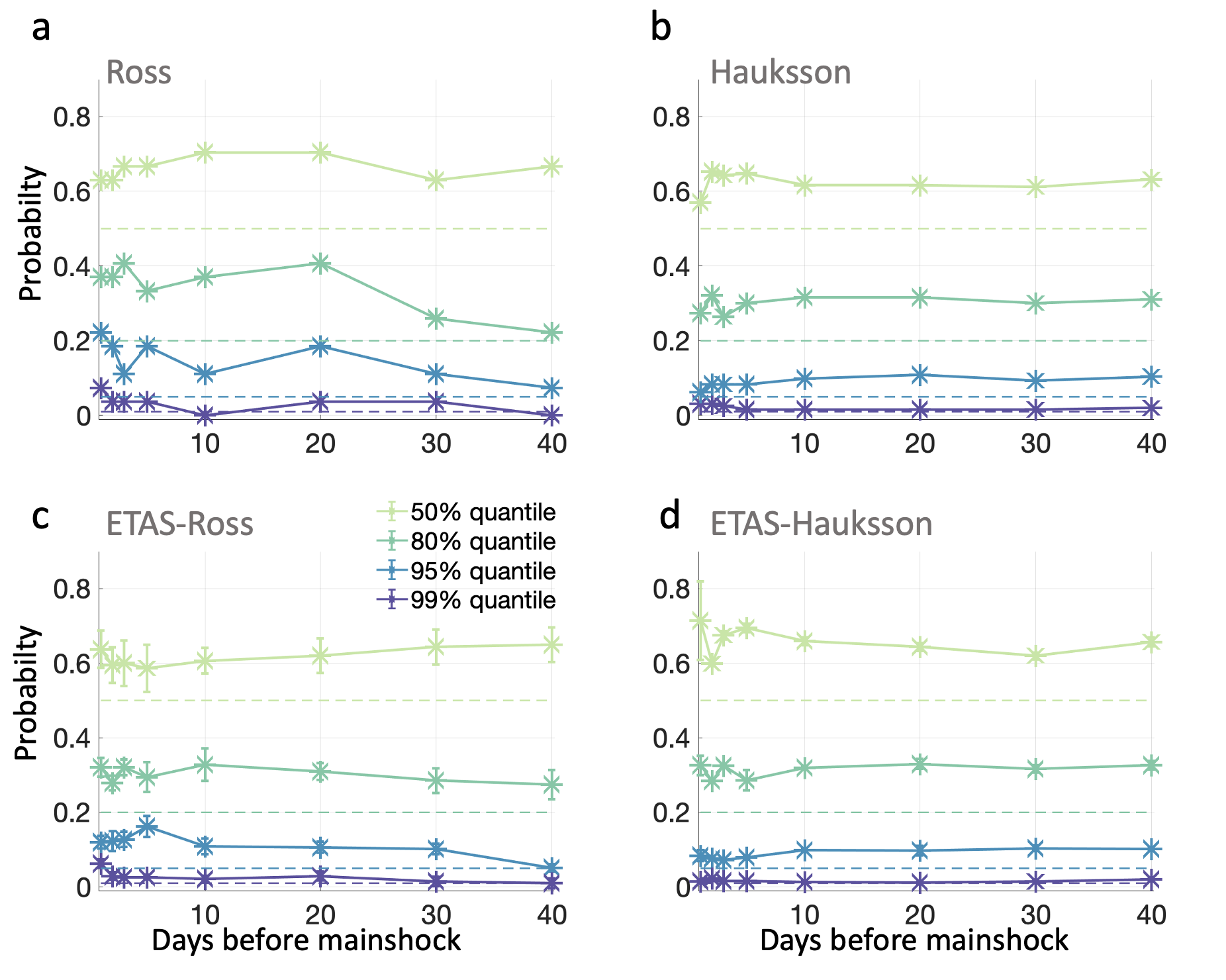}
	\caption{
        {\bf Quantiles for foreshock flag=1. } Same as Fig. \ref{fig:Fig3}, with foreshock flag = 1.}  
	\label{fig:SI_Fig1_Omori}
\end{figure}

\begin{figure}[ht!]
\centering
	\includegraphics[width=\linewidth]{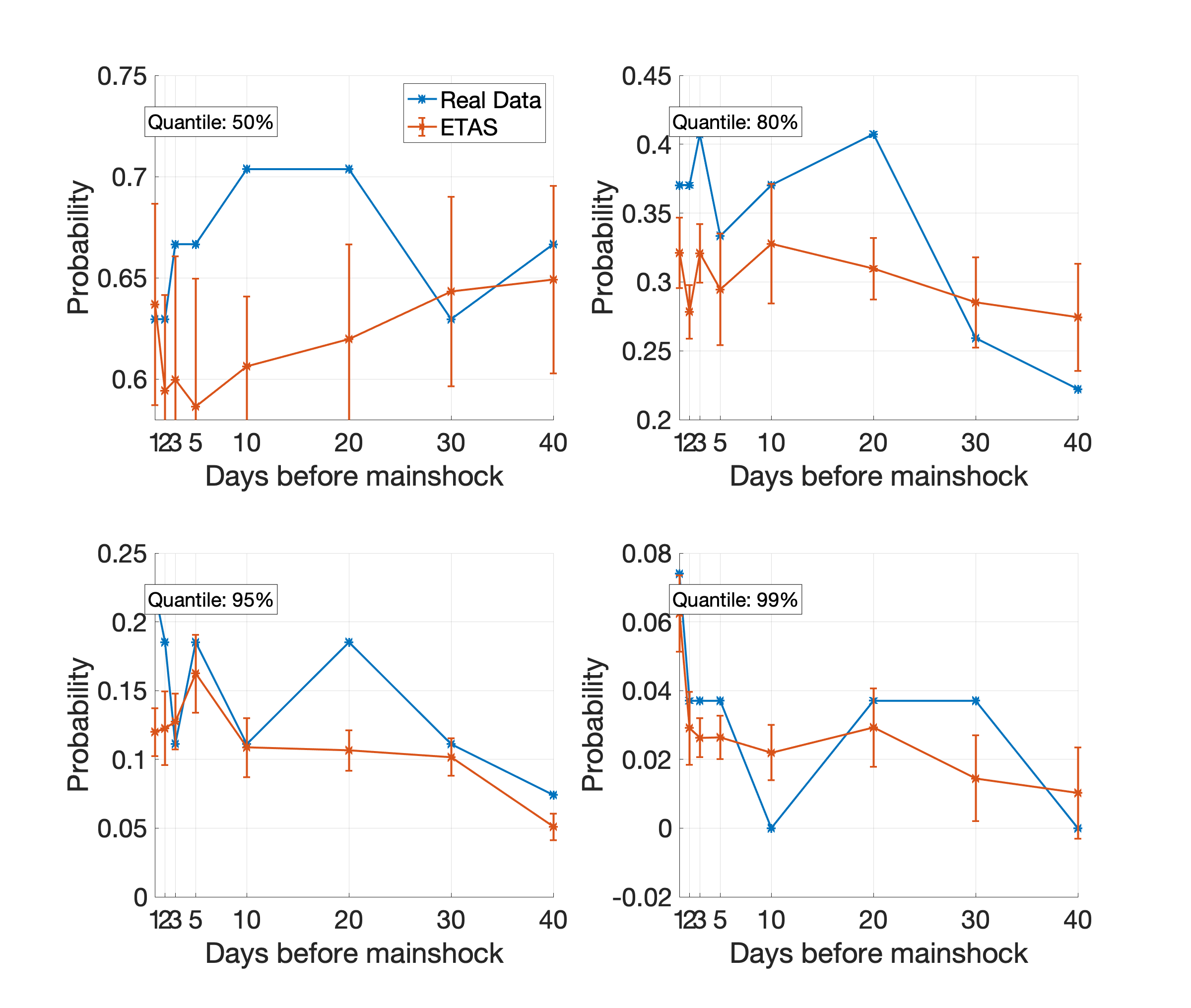}
	\caption{
        {\bf Detailed comparison of the quantiles results the Ross-Ross ETAS catalogs.} Magnified version of Fig.~\ref{fig:Fig3}(a,c) for the Ross {\it et al.} catalog. foreshock flag = 1.}  
	\label{fig:SI_Fig8_RossETAS_comp}
\end{figure}

\begin{figure}[ht!]
\centering
	\includegraphics[width=\linewidth]{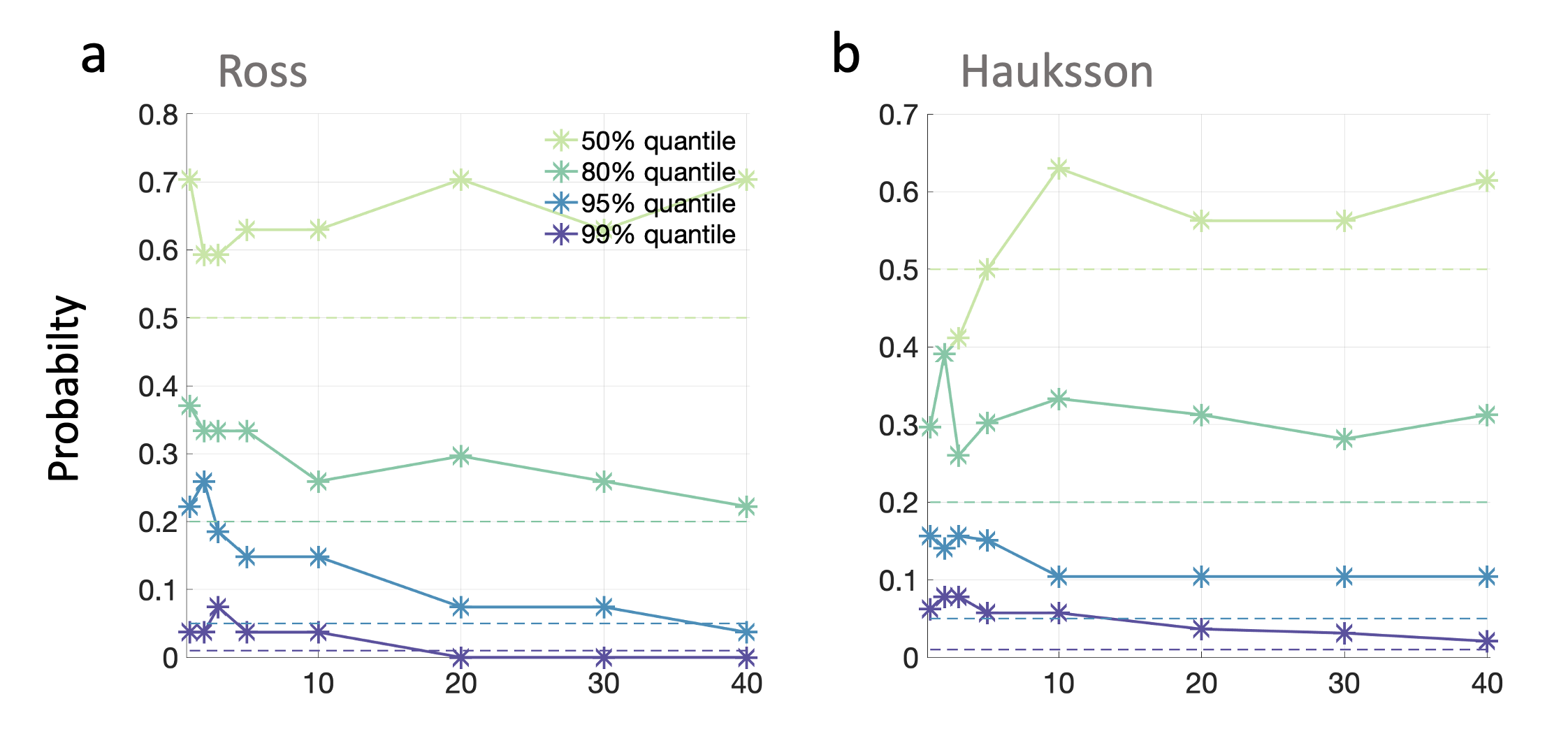} 
	\caption{
    {\bf Quantiles when imposing a distance constraint.} Same as Fig. \ref{fig:Fig3} but imposing a distance constraint of 100 km from the epicenter. Here, the magnitude threshold is 4.5, and the foreshock flag = 0. }
	\label{fig:SI9}
\end{figure}
\end{document}